\documentclass{elsart}
\usepackage{epsfig}
\usepackage{amsmath}
\usepackage{amsfonts}
\usepackage[usenames]{color}
\usepackage{graphicx,slashbox}
\usepackage{ulem}

\newcommand{\JMcomm}[1]{{\color{black} #1}}

\newcommand{\MDcomm}[1]{{\color{black} #1}}

\newcommand{\JMcommNEW}[1]{{\color{black} #1}}  %red
\newcommand{\MDcommNEW}[1]{{\color{black} #1}}  %OliveGreen
\newcommand{\EScommNEW}[1]{{\color{black} #1}}  %blue

\begin{document}

\begin{frontmatter}

%\title{Generalized SIC-Slater and static polarizabilities.}
\title{Polarizibilities as a test of localized approximations to the
  self-interaction correction} 

\author{J. Messud\corauthref{cor}$^{a,b}$,}
\author{Z. Wang$^{a,b,c}$,}
\author{P.~M.~Dinh$^{a,b}$,}
\author{P.-G.~Reinhard$^d$ and E.~Suraud$^{a,b}$}

\corauth[cor]{Corresponding author\\{\it Email-address}~:
  messud@irsamc.ups-tlse.fr} 
\address{$^a$Universit\'e de Toulouse; UPS; \\Laboratoire de Physique
  Th\'eorique (IRSAMC); F-31062 Toulouse, France}
\address{$^b$ CNRS; LPT (IRSAMC); F-31062 Toulouse, France}
\address{$^c$ Institute of Low energy Nuclear Physics, 
Beijing Normal University, Beijing 100875, China}
\address{$^d$Institut f{\"u}r Theoretische Physik, Universit{\"a}t
  Erlangen, D-91058 Erlangen, Germany}

\begin{abstract}
We present applications of the recently introduced ``Generalized
SIC-Slater'' scheme which provides a simple Self-Interaction
Correction approximation in the framework of the Optimized Effective
Potential. We focus on the computation of static polarizabilities
which are  known to constitute stringent tests for Density Functional
Theory.
%Within the Self Interaction Correction (SIC) formalism applied to the Density Functional Theory
%we study the properties of the recently proposed ``Generalized SIC-Slater'' scheme. 
%We discuss 3D numerical results on an observable which is especially 
%sensitive to details of the SIC, ie polarization.
We apply the new method to model H chains, but also to more
realistic systems such as C$_4$ (organic) chains, and 
\EScommNEW{less symmetrical systems such as a } Na$_5$ (metallic)
cluster.
%% and H$_4$ ground state (``T-shaped'')
%% configuration. 
Comparison is made with other SIC schemes, especially
with the standard SIC-Slater one. 
\end{abstract}


\end{frontmatter}

\section{Introduction}

{Density-functional theory (DFT) has become over the years 
one of the most powerful theories for the description 
of complex electronic systems ranging from atoms and molecules, to bulk solids. It allows realistic 
calculations of an ever increasing number of systems in physics 
and chemistry \cite{Koh99aR,Par89B,Dre90B}.}
%
%It is one of the only theory which allows to
%compute complex systems, because of it's low numerical cost \cite{Par89B,Dre90B}. 
As the exact functional is not known, most applications employ the Local Density Approximation (LDA),
see e.g. \cite{Jon89aR}, 
or its extension to the Generalized Gradient
Approximation (GGA) \cite{Per96a}. In spite of their successes, these
approaches still have deficiencies. In particular, the
self-interaction error spoils single-particle properties as, e.g., the
ionization potential % or the band gap in solids
\cite{Hyb86a,Nie00aR}. 
Another critical detail where LDA and GGA {usually} fail
is the polarizability in chain molecules \cite{Gis99a,Kue04a}.
An intuitive and efficient solution
%, close to the spirit of DFT, 
is to augment LDA by a
Self-Interaction Correction (SIC) \cite{Per79a,Per81a}, i.e. to 
{introduce an explicit orbital dependence of the functional by subtracting 
by hand the spurious self-interaction.}
%deal with single density functionals. 
The drawback is that it
produces a state-dependent mean-field Hamiltonian which
requires extra efforts to {enforce} orthogonality of the single
particle basis \cite{Per81a,Har83a,Goe97a,Mes08-2}.
The optimized effective potential (OEP) method
\cite{Kue07aR,OEP1,OEP2,TDOEP}
overcomes that complication as it allows
%permits 
to define the best common (state-independent) local mean-field
potential $V(\mathbf{r})$.  
%% Indeed it is found that OEP manages to maintain crucial features of
%% the underlying SIC as, e.g., the localization of states or the
%% derivative discontinuity \cite{Per83a,Kri90a,Mun05a},
\MDcommNEW{Indeed some crucial features of
the underlying SIC as, e.g., the localization of states or the
derivative discontinuity \cite{Per83a,Mun05a} are found to be
maintained through an OEP procedure~\cite{Kri90a}, for a comprehensive
overview see \cite{Kue07aR}.}
But the exhaustive SIC-OEP equations are difficult to handle and are thus
often simplified. {A most} popular approximation is the so-called
%is, e.g., the 
Krieger-Li-Iafrate (SIC-KLI)
approach \cite{OEP1,OEP2} and, in a further step of simplification,
the SIC-Slater approximation \cite{Sha53a}. 
However, SIC-KLI and SIC-Slater
approximations can easily miss crucial features of SIC as, e.g., 
%the localization of {orbitals} 
%states and 
the performance with respect to polarizability, even if accurate
exchange-correlation potentials are used \cite{Kor08}
(note that the same conclusion holds at the exact-exchange level \cite{Kue04a}).

%And even when a full OEP resolution is done,
%the role of the localization and the single electron interpretation are not fully clear.

A key question is the localization of orbitals, as had been
already observed rather early \cite{Edm63aR}, which
then minimizes the electronic interaction energy.
Indeed states which optimize SIC tend to be localized whereas states
emerging from a common local mean-field tend to \EScommNEW{more}
delocalization. These 
contradicting demands can be bridged by full OEP but spoil approximate
treatments as SIC-KLI or SIC-Slater approximation. A fairly convenient way out
is to use two sets of orbitals. That was already proposed in
\cite{Ped84} and has been used to improve on the SIC-KLI approximation
\cite{Gar00,Pat01,Pem08}, although not yet in a fully consistent and
variational form. Taking up the double-set idea, we have recently
proposed a SIC-OEP scheme relying on two sets of {complementing}
orbitals \cite{Mes08-1}.
At the purely stationary \EScommNEW{full} SIC level, it is mostly a matter of
convenience, \MDcommNEW{while} a double-set technique becomes crucial in
practical applications of the time-dependent SIC \cite{Mes08-2}. 
\MDcommNEW{In this paper, we will also demonstrate that this method can be
powerfully exploited in stationary calculations in the frame of SIC-OEP. Indeed}
the two sets are connected by a unitary transformation, thus building
the same total density. One of the sets remains spatially localized
while the other set is free to accommodate a common local
potential and/or minimal energy variance. The spatially localized set
is determined by variation of the SIC energy with respect to unitary
transformation coefficients which allows that these states fulfill
what is called the ``symmetry condition''
\cite{Mes08-2,Ped84,Mes08-1,Mes09}.
%When one considers OEP approximations to SIC the double set picture 
%allows to clarifies the role of localization.
The localized set shows a much better performance with
respect to \MDcommNEW{standard} SIC-KLI or SIC-Slater approximation. 
\MDcommNEW{The resulting formalism is called ``Generalized SIC-OEP'' \cite{Mes08-1}.
We further simplified the resulting equations,
following the track of the SIC-Slater approximation and developed in a
strictly variational manner a double-set treatment of the SIC-Slater
approximation to OEP, namely the ``Generalized SIC-Slater''
approximation.}
It is to be noted that a very similar
development is found in \cite{Kor08-2}.  As one of the sets remains
spatially localized, it validates the Generalized SIC-Slater
approximation to Generalized SIC-OEP built from this set, while
maintaining key features of the full SIC scheme: it is energetically
advantageous for the SIC energy, permits to re-establish potential energy surfaces (PES), and performs fairly
well in the calculations of polarizabilities. 
We pointed out that the Generalized SIC-Slater emerges
naturally because of the localization of one set of orbitals
\cite{Mes08-1}, whereas K\"orzd\"orfer \textit{et al.} \cite{Kor08-2}
\MDcommNEW{compared
Generalized SIC-KLI and full Generalized SIC-OEP with standard SIC-KLI and
OEP and also with the approximate Generalized SIC-KLI of \cite{Pem08}.}
%worked at the level of the SIC-KLI approximation.  
\JMcomm{It is to be noted that}
\EScommNEW{the practical introduction of} 
%previous attempts to introduce ``by hand'' 
localized orbitals within
the SIC-KLI approximation \EScommNEW{has also been proposed in}
\cite{Gar00,Pat01,Pem08} \EScommNEW{without explicit ``symmetry condition''}. 
%none of them introduced the crucial ``symmetry condition'', and the
The form of the SIC-KLI approximation used in those papers is not exactly what comes 
out from the more fundamental ``Generalized SIC-OEP''
\cite{Kor08-2}.
%We called the new scheme ``generalized SIC-Slater''.
After a brief presentation of the formalism, we apply it to model
hydrogen % H$_4$ 
chains, and to various other 
%more realistic 
systems such as C (organic) chains and  Na (metallic) clusters. 
% and H$_4$ ground state (``T-shaped'')
%configuration.  
%%Comparisons are  made with other SIC schemes, especially
%with the usual SIC-Slater one.

\section{The Generalized SIC-Slater formalism}

\subsection{Summary of SIC equations}
\label{sec:formalism}

We briefly summarize the formalism using for simplicity
a notation without explicit spin densities. The generalization to
these is obvious. The calculations later on use, of course, the full
spin density functional.

The starting point for {the formulation of} SIC 
%considerations 
is the SIC energy functional for electrons
\begin{eqnarray}
  E_\mathrm{SIC}
  =
  E_\mathrm{kin}
  \!+\!
  E_\mathrm{ion}
  \!+\!
  E_\mathrm{\rm LDA}[\rho]
  \!-\!
  \sum_{\beta=1}^{N} E_\mathrm{\rm LDA}[\rho_{\beta}]
  \;,\;
  \rho
  =
  \sum_{\beta=1}^{N}\rho_{\beta}
  \;,\;
  \rho_{\beta}
  =
  |\psi_\beta|^2
\label{eq:fsicen}
\end{eqnarray}
where $E_\mathrm{\rm LDA}[\rho]$ is a standard LDA energy-density functional
\JMcomm{(which contains both the Hartree and the exchange-correlation energies in our notations)}
complementing the kinetic energy $E_\mathrm{kin}$ and the
interaction energy with the ionic background $E_\mathrm{ion}$. The
last term is the SIC correction.
{The densities $\rho_\alpha$ and $\rho$ are defined from the
 set of occupied single-particle states $\{\psi_\beta,\beta=1...N\}$.}
The SIC equations are obtained by standard variational techniques within imposing explicit 
orthonormalization of the orbitals by a set of Lagrange multipliers $\lambda_{\alpha\beta}$.
We now introduce a second set of orbitals $\{\varphi_i\}$
related to the previous one by a unitary
transformation within the set of occupied states
(i.e. leading to the same total density $\rho$~: $\varphi_i=\sum_{\alpha}u^*_{i\alpha} \psi_\alpha$)
which diagonalizes the $\lambda_{\alpha\beta}$.
%SIC Hamiltonian
We can then recast the resulting equations in eigenvalue equations \cite{Ped84,Mes08-1,Mes09} 
\begin{eqnarray}
\hat{h}_\mathrm{SIC}|\varphi_i)
  &=&
  \varepsilon_i |\varphi_i)
\label{eq:static-diaq}
\\
0 &=&  (\psi_\beta|U_\beta-U_\alpha|\psi_\alpha)
\label{eq:symcond2} 
\end{eqnarray}
with the SIC Hamiltonian reading
\allowdisplaybreaks
\begin{subequations}
\label{eq:SICmfham}
\begin{eqnarray}
  \hat{h}_\mathrm{SIC}
  &=&
  \hat{h}_\mathrm{\rm LDA}
  -
  \sum_\alpha U_\alpha|\psi_\alpha)(\psi_\alpha|
\label{eq:hsic}
%\\
\\
  \hat{h}_\mathrm{\rm LDA}
  &=&
  \frac{\hat{p}^2}{2m}
  +
 U_{\rm LDA}\left[\rho\right](\mathbf{r})
  , \quad {\rm with} \quad
  U_{\rm LDA}\left[\rho\right](\mathbf{r})
  =
  \frac{\delta E_\mathrm{\rm LDA}[\rho]}
       {\delta \rho(\mathbf{r})}
\label{eq:SICmf}
\\
  U_\alpha(\mathbf{r})
  &=&
  \frac{\delta E_\mathrm{\rm LDA}[\rho_\alpha]}
       {\delta \rho_\alpha(\mathbf{r})}
  =
  U_{\rm LDA}\left[ |\psi_\alpha|^2 \right](\mathbf{r})
\end{eqnarray}
\end{subequations}
where $\hat{h}_\mathrm{\rm LDA}$ is the standard LDA Hamiltonian.
%and the second term in Eq. (\ref{eq:hsic}) 
%comes from the SIC term in the energy (\ref{eq:fsicen}).
%And the variational principle jointly leads to the  crucial  
%complement to 
%Eq. (\ref{eq:static-SIC}), that is, to the 
%symmetry condition  Eq.(\ref{eq:symcond2}).
%\begin{eqnarray}
%  0
%  =
%  (\psi_\beta|U_\beta-U_\alpha|\psi_\alpha)
%  \quad.
%\label{eq:symcond2}
%\end{eqnarray}
%
%\begin{eqnarray}
%  && 
%\psi_\alpha
%  =
%  \sum_{i} \varphi_i \, u_{i\alpha}
%\label{eq:unitrans}
%\end{eqnarray}
%
The coefficients $u_{i\alpha}$ of the unitary transformation
for given \JMcomm{\textit{diagonal} orbitals} $\varphi_i$ are determined such
that the $\psi_\alpha$ satisfy the symmetry condition
(\ref{eq:symcond2}). \JMcomm{The $\psi_\alpha$ are called \textit{localized} orbitals because they are spatially much more localized \cite{Ped84,Mes09}.}

\subsection{SIC and OEP}

The eigenvalue equation (\ref{eq:static-diaq}) employs a non-local
Hamiltonian $\hat{h}_{\rm SIC}$, 
see Eq.~(\ref{eq:hsic}), which complicates the numerical handling.
In \cite{Mes08-1}, we proposed to apply the OEP formalism to this two-sets SIC formulation,
to find the best local approximation to its Hamiltonian. 
\JMcomm{A very similar development is found in \cite{Kor08-2}.}
We start from a set $\varphi_i$ 
\JMcomm{which satisfies the eigenvalue equations}
(this set is not exactly the same as that of the exact SIC
equation because additional restriction of the Hilbert space is imposed here --
this point being clarified, we will employ the same symbol to simplify the notations)~:
\begin{eqnarray}
  \left[\hat{h}_\mathrm{LDA}({\bf r})-V_0(\mathbf{r})\right] \varphi_i({\bf r})
  =
  \varepsilon_i \varphi_i({\bf r})
  \quad,
\label{eq:eigen3}
\end{eqnarray}
where $V_0$ is a local and state-independent potential which needs to
be optimized to minimize the SIC energy~(\ref{eq:fsicen}). It is
important to note that this energy is still expressed in
terms of the $\psi_\alpha$, linked by a unitary transformation
to the $\varphi_i$ and which satisfy the
symmetry condition (\ref{eq:symcond2}) in our case.  The optimized
effective potential $V_0(\mathbf{r})$ is found by variation $\delta
E_{\rm SIC} / \delta V_0(\mathbf{r})=0$.
\JMcommNEW{
We obtain
$
V_0 = V_{\rm S}+V_{\rm K}+V_{\rm C}
$,
with \cite{Mes08-1,Kor08-2}
\begin{subequations}
\begin{eqnarray}
V_{\rm S} &=& \sum_\alpha \frac{|\psi_\alpha|^2}{\rho} U_\mathrm{\rm
  LDA}[|\psi_\alpha|^2]\quad, 
\label{eq:pot_slat10} \\
V_{\rm K} &=& \frac{1}{\rho} \sum_{\alpha,\beta} \Big( \sum_i
|\varphi_i|^2 \upsilon_{i\alpha}^{*}\upsilon_{i\beta} \Big) ({
  \psi_\beta}|V_0-U_\mathrm{\rm LDA} [|\psi_\alpha|^2]|\psi_\alpha) \quad,
\label{eq:pot_SIC-KLI10} \\
V_{\rm C} &=& \frac{1}{2}\sum_i \frac{\mathbf{\nabla}.(p_i
  \mathbf{\nabla}|\varphi_i|^2)}{\rho} \quad,
\label{eq:pot_OEP10} \\
p_i(\mathbf{r})&=&\frac{1}{\varphi_i^*(\mathbf{r})}
\sum_\alpha\!\upsilon_{i\alpha}\int\!\!\textrm d\mathbf{r'} \Big(
V_0(\mathbf{r'})-U_\mathrm{\rm LDA} [|\psi_\alpha|^2](\mathbf{r'})
\Big) \psi_\alpha^*(\mathbf{r'})G_i(\mathbf{r},\mathbf{r'}) ,
\label{eq:p_i10}\\
G_i(\mathbf r,\mathbf r') &=& \sum_{j\neq i}
\frac{\varphi_j^*(\mathbf r) \varphi_j(\mathbf r')} {\varepsilon_j -
  \varepsilon_i} \quad.
\label{eq:G_i10}
\end{eqnarray}
\end{subequations}
}

\subsection{Generalized SIC-Slater}

The involved OEP
equations can be simplified by exploiting the property that the
$\psi_\alpha$ are spatially localized \cite{Ped84},
\JMcommNEW{ which yields $V_0|\psi_\alpha)\approx 
U_{\rm LDA} [|\psi_\alpha|^2]|\psi_\alpha)$.}
This allows to employ the SIC-Slater approximation to OEP, yielding
\cite{Mes08-1}~:
\begin{eqnarray}
  V_0(\mathbf{r})
  &\simeq&
  \sum_\alpha \frac{|\psi_\alpha(\mathbf{r})|^2}{\rho(\mathbf{r})}
  U_{\rm LDA}[|\psi_\alpha|^2](\mathbf{r})
  \quad,
\label{eq:SIC-Slater3}
\end{eqnarray}
Note that this equation has the form of a SIC-Slater approximation
\cite{Kri90a,Slat51} but is constructed from the \JMcomm{localized} orbitals
$\psi_\alpha$ and is applied to the \JMcomm{diagonal orbitals} $\varphi_i$. We called this new
scheme ``Generalized SIC-Slater''(GSlat)  approximation, which differs from
the standard SIC-Slater scheme because of the two basis sets involved
here and which, therefore, has 
\JMcomm{more flexibility.}
%a broader range of validity.
%
The practical scheme for GSlat can be summarized as
follows~: {\it i)} Eq.~(\ref{eq:eigen3}) generates
the ``diagonal'' set $\varphi_i$ of occupied states; {\it ii)} the
unitary transformation serves to accommodate the
symmetry condition (\ref{eq:symcond2}) which, in turn, defines the
``localized'' set $\psi_\alpha$; {\it iii)} the latter set enters the
OEP $V_0$ as given in Eq.~(\ref{eq:SIC-Slater3}). 

\subsection{Generalized SIC-KLI and approximations thereof}

\JMcommNEW{
Even if the GSlat approximation \MDcommNEW{might} be satisfying in
most cases, it is worth discussing in more detail the SIC-KLI correction
(\ref{eq:pot_SIC-KLI10}). 
The use of the localization of the $\psi_\alpha$ allows to keep only
the diagonal terms $\alpha=\beta$, that is,
\begin{eqnarray}
V_{\rm K}\approx \frac{1}{\rho} \sum_{\alpha} \Big( \sum_i
|\upsilon^*_{i\alpha}\varphi_i|^2 \Big) ({
  \psi_\alpha}|V_0-U_\mathrm{\rm LDA} [|\psi_\alpha|^2]|\psi_\alpha). 
\label{eq:pot_OEP4}
\end{eqnarray}
In contrast to the GSlat approximation, fluctuations of
$V_0|\psi_\alpha)$ around $U_{\rm LDA}
[|\psi_\alpha|^2]|\psi_\alpha)$ are not neglected, even if one expects
them to remain small.}

\MDcommNEW{It is now instructive to use the following identity~:
\[
|\sum_i \upsilon^*_{i\alpha}\varphi_i|^2 = \sum_i
|\upsilon^*_{i\alpha}\varphi_i|^2 
+ \sum_{\stackrel{i,j}{i \ne j}} (\upsilon^*_{i\alpha}
\varphi_i)(\upsilon_{j\alpha}^* \varphi_j)^*,
\]
to multiply it by $({ \psi_\alpha}|V_0-U_\mathrm{\rm LDA}
[|\psi_\alpha|^2]|\psi_\alpha)$ and to sum over $\alpha$. We thus
obtain~:
\begin{subequations}
\begin{eqnarray}
%&&
\sum_{\alpha} |\psi_\alpha|^2 ({ \psi_\alpha}|V_0-U_\mathrm{\rm LDA}
    [|\psi_\alpha|^2]|\psi_\alpha)
%\nonumber\\
%&& \qquad = \sum_i |\varphi_i|^2 \sum_{\alpha} |\upsilon_{i\alpha}|^2 ({
%      \psi_\alpha}|V_0-U_\mathrm{\rm LDA}
%    [|\psi_\alpha|^2]|\psi_\alpha) 
%\label{eq:A_ii}\\
%&&\qquad \qquad+ \sum_{\stackrel{i,j}{i \ne j}} \varphi_i\varphi_j^*
%\sum_{\alpha}\upsilon^*_{i\alpha}\upsilon_{j\alpha} ({
%  \psi_\alpha}|V_0-U_\mathrm{\rm LDA} [|\psi_\alpha|^2]|\psi_\alpha) 
%\label{eq:A_ij} \quad,\\
%&& 
%\qquad 
= \sum_i A_{ii} + \sum_{\stackrel{i,j}{i \ne j}} A_{ij} \quad
%\simeq \quad V_{\rm K} + \sum_{\stackrel{i,j}{i \ne j}} A_{ij}
\quad, 
\label{eq:A_ii&A_ij}
\end{eqnarray}
\end{subequations}
where 
%we have used in the last equation the approximate form (\ref{eq:pot_OEP4}) of the SIC-KLI correction and defined the matrix element 
\begin{eqnarray}
A_{ij} = \varphi_i\varphi_j^* 
\sum_{\alpha}\upsilon^*_{i\alpha}\upsilon_{j\alpha} ({
  \psi_\alpha}|V_0-U_\mathrm{\rm LDA} [|\psi_\alpha|^2]|\psi_\alpha)
\label{eq:A_ij}
.
\end{eqnarray}
\JMcommNEW{First, we have the expression $V_{\rm K} \approx
\frac{1}{\rho}\sum_i A_{ii}$ where we used the approximate form
(\ref{eq:pot_OEP4}) of the SIC-KLI correction. 
If one \MDcommNEW{further} assumes that for $i\neq j$, $|A_{ij}| \ll
|A_{ii}|$, one obtains the following approximation of the SIC-KLI
correction
\begin{eqnarray}
V_{\rm K} \approx \frac{1}{\rho} \sum_{\alpha} |\psi_\alpha|^2 ({
  \psi_\alpha}|V_0-U_\mathrm{\rm LDA} [|\psi_\alpha|^2]|\psi_\alpha)
\quad,
\label{eq:pot_OEP5}
\end{eqnarray}
which is the form proposed in \cite{Gar00,Pat01,Pem08}
without reference to the symmetry condition. This expression can
thus be derived from an approximation of the so-called ``Generalized
SIC-KLI'' potential. We will call it ``Localized SIC-KLI'' (Loc. SIC-KLI) thereafter.}

%\JMcommNEW{
%A way to justify that for $i\neq j$, $A_{ij} \ll A_{ii}$ is to assume
%that all terms $({ \psi_\alpha}|V_0-U_\mathrm{\rm
%    LDA} [|\psi_\alpha|^2]|\psi_\alpha)$ take about the same (small)
%value (i.e. that their value is almost independent of
%$\alpha$.), then $A_{ij}$ reduces to
%\[
%A_{ij} \propto \varphi_i\varphi_j^*  \sum_{\alpha} \upsilon^*_{i\alpha}
%\upsilon_{j\alpha}= \varphi_i\varphi_j^* \delta_{ij},
%\]
%since $\hat{\upsilon}$ is a unitary transformation. These assumption probably holds
%}
%One finally gets
%the following approximation of the SIC-KLI correction~:
%\begin{eqnarray}
%V_{\rm K} \approx \frac{1}{\rho} \sum_{\alpha} |\psi_\alpha|^2 ({
%  \psi_\alpha}|V_0-U_\mathrm{\rm LDA} [|\psi_\alpha|^2]|\psi_\alpha)
%\quad,
%\label{eq:pot_OEP5}
%\end{eqnarray}
%
%which is the form proposed in \cite{Gar00,Pat01,Pem08}
%without reference to the the symmetry condition. This expression can
%thus be derived from an approximation of the so-called ``Generalized SIC-KLI''
%based on the property that $({\psi_\alpha}|V_0-U_\mathrm{\rm
%  LDA} [|\psi_\alpha|^2]|\psi_\alpha)$ is almost independent of
%$\alpha$.
%We can also derive Eq.~(\ref{eq:pot_OEP5}) with a less
%restrictive assumption, that is, for $i\neq j$, $A_{ij} \ll A_{ii}$,
%which leads to the same approximation.
%These assumptions probably hold
%for spatially symmetric systems, as will be discussed lengthly when
%tested on Hydrogen chains (see Sec.~\ref{sec:Hchains}). 
The justification of the approximation $|A_{ij}| \ll |A_{ii}|$ remains 
to be clarified. A way to circumvent the formal difficulty is to 
consider practical applications to see how the approximation performs. Still, 
the assumption probably holds
for spatially symmetric systems, as will be discussed lengthly when
tested on hydrogen chains (see Sec.~\ref{sec:Hchains}). The case of 
asymmetrical systems however  remains to be explored in more detail, 
%an open question how Eq.~(\ref{eq:pot_OEP5}) performs, 
especially in comparison to the
formally better founded forms (\ref{eq:pot_SIC-KLI10}) or
(\ref{eq:pot_OEP4}).
}

\subsection{Summary}

\MDcommNEW{All SIC mean-field Hamiltonians presented above enter a
Schr\"odinger-like equation of the form
$\hat{h}|\varphi_i)=\epsilon_i|\varphi_i)$. We have thus summarized
them in table \ref{tab:schemes}, so that one can easily track the
various contributions from one Hamiltonian to another.
%% In the following, we compare the results for 
%% GSlat with full SIC and other approaches. The corresponding
%% mean-field Hamiltonians are summarized in table \ref{tab:schemes},
%
%% all being used in one-body eigenvalue equations of the form 
%% $\hat{h}|\varphi_i)=\epsilon_i|\varphi_i)$.
%
\begin{table}[htbp]
\begin{center}
\begin{tabular}{|l|l|}
\hline
Expression of $\hat{h}$ in $\hat{h}|\varphi_i)=\epsilon_i|\varphi_i)$&
Method\\ 
\hline
\hline
$\hat{h}_{\rm LDA}[\rho]$ & LDA \\
\hline
$\displaystyle
\hat{h}_{\rm LDA}[\rho]-\hat{U}_{\rm LDA} \left[ \frac{\rho}{N}
  \right]$ & Average Density SIC \\ 
\hline
$\displaystyle
\hat{h}_{\rm LDA}[\rho]-\sum_j \frac{|\varphi_j|^2}{\rho} \hat{U}_{\rm
  LDA}\left[|\varphi_j|^2\right]$ & Standard SIC-Slater \\ 
\hline
$\displaystyle
\hat{h}_{\rm LDA}[\rho] - \sum_\alpha \frac{|\psi_\alpha|^2}{\rho}
\hat{U}_{\rm LDA}\left[|\psi_\alpha|^2\right]$ & Generalized
SIC-Slater \\   
\hline
\MDcommNEW{$\displaystyle
\hat{h}_{\rm LDA}[\rho] - \sum_\alpha \frac{|\psi_\alpha|^2}{\rho}
\hat{U}_{\rm LDA}\left[|\psi_\alpha|^2\right]$} & \JMcommNEW{Localized SIC-KLI}
\\ 
\hspace{1.25cm} 
\MDcommNEW{$\displaystyle
-  \frac{1}{\rho} \sum_{\alpha} |\psi_\alpha|^2 ({
  \psi_\alpha}|V_0-U_\mathrm{\rm LDA} [|\psi_\alpha|^2]|\psi_\alpha)$}
& \\ 
\hline
$\hat{h}_{\rm LDA}[\rho] - \sum_\alpha \hat{U}_{\rm
  LDA}\left[|\psi_\alpha|^2\right] |\psi_\alpha)(\psi_\alpha|$ &
  Exact SIC (benchmark)\\ 
\hline
\end{tabular}
\caption{\label{tab:schemes}
The hierarchy of mean-field Hamiltonians, from simple-most LDA (top
line) to full SIC (bottom line), called ``exact'' SIC.
}
\end{center}
\end{table}
Note that the symmetry condition (\ref{eq:symcond2}) should be added
for the last three schemes, to define the localized states $\psi_\alpha$
{required in} the corresponding Hamiltonians.
}

\subsection{Computational details}

The test cases presented in this paper have been obtained using a full
3D DFT code originally developed for large scale calculations of the
dynamics of metal clusters \cite{Cal00} and later on small hydrogen
clusters \cite{Ma05}, and now extended to treat any organic system
\cite{Wan09}. The electronic wave functions are represented on an
equidistant 3D grid with fixed mesh size
\JMcommNEW{(between 0.4 and 0.8 $a_0$, depending on the studied system)}. 
The ionic background is
treated by means of pseudopotentials, using either local ones (Na, H)
\cite{Kue98} or non-local ones to treat organic systems
\cite{Goe96}.  Even in the case of hydrogen, we use a pseudopotential
in order to regularize the Coulomb singularity at origin (\JMcomm{the
Giannozzi pseudopotential as in \cite{Kor08}}).
The pseudopotential
parameters, especially the core size, fixes the optimal grid
representation.  \JMcomm{For the molecular chains calculations, we use
boxes of $40-52$ $a_0$ in the longitudinal direction (according to the
case) and 20 $a_0$ in the transverse directions. For the Na$_5$
cluster, we use a box of 38 $a_0$ in each direction.}  Those sizes are
sufficiently large to provide good convergence properties of
calculations. For completeness, computations have been checked using
larger boxes with no noticeable differences in the results.
Electronic ground states are obtained by damped gradient iterations.
%% \footnote{ES2o: referee II ask for consistency of pseudo with functional... Should we enter
%% such details or try to overlook the case ?
%% \\
%% \PGRcomm{PGR2o: This is opening a nest of worms. I would confine a
%%   statement to the reply.}}

\section{Static polarizability results}

We had shown in \cite{Mes08-1} that GSlat solves the
problem {with potential-energy surfaces} encountered in the
standard SIC-Slater scheme and {produces good results for
the polarizability} of the C atom.
Here we {will} show through full 3D calculations that it also
{yields favorable results} for more complex structures: model
H chains, C$_4$ chains and  a Na$_5$ cluster. 
% and the ``T-shaped''
%H$_4$ ground state configuration.  
We compare the GSlat results to LDA, ADSIC (Average Density SIC)
\cite{Leg02}, standard SIC-Slater and exact SIC results, the latter being
the benchmark.
{For the comparison, we use the static dipole polarizability as
a most} sensitive test for DFT approximations \cite{Gri01a}.
Considering a system put inside an electrical field ${\bf E}$,
{the polarizability} is defined as $\alpha_i=\partial \mu_i /
\partial E_i$, where $\mu_i$ is the dipole moment along the $i$
direction and $E_i$ the electric field along $i$.

\subsection{Hydrogen chains}
\label{sec:Hchains}

Linear chains of H atoms constitute (highly) simplified model systems
for various important chain or chain-like molecules such as in
particular poly-acetylene with its remarkable properties
\cite{Chi77}.
These model systems are of great interest to
investigate DFT schemes \cite{Kor08,Kue04} as they are particularly
difficult to be correctly described within {\rm LDA} \cite{Pat01}.  
They thus provide a critical test.  

\begin{figure}[htbp]
\centerline{\includegraphics[width=0.95\linewidth,angle=0]{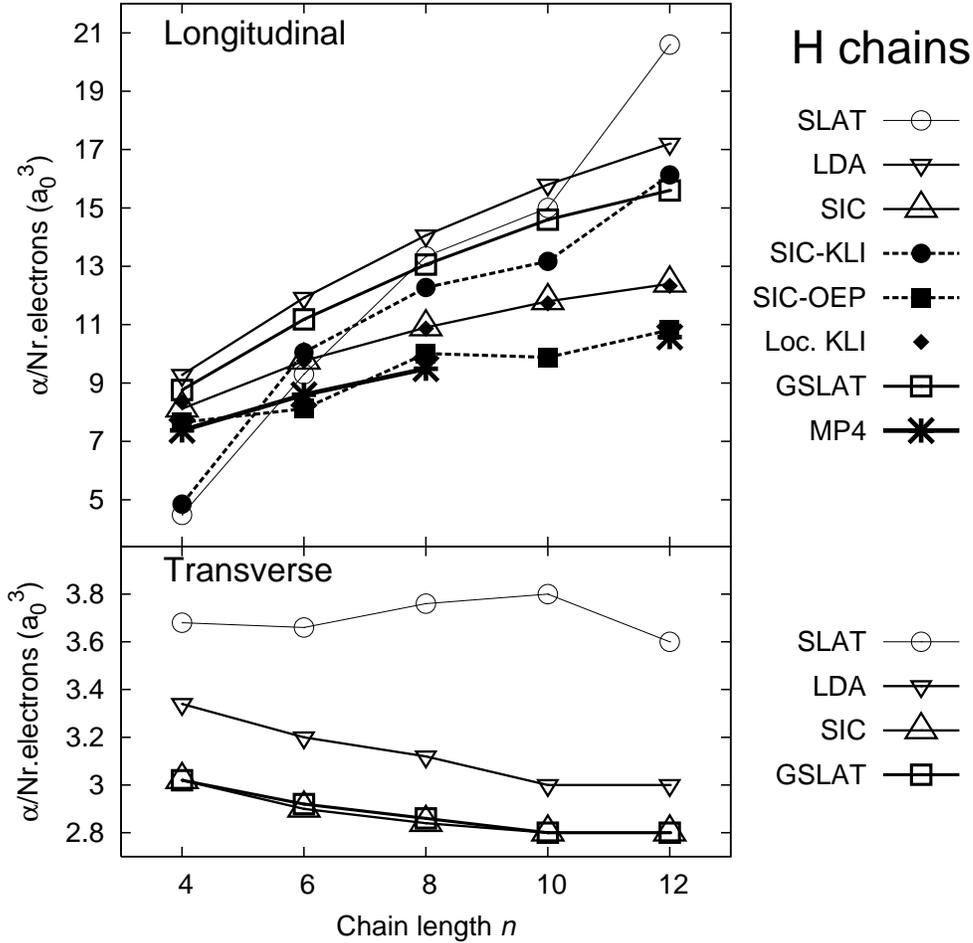}}
\caption{\label{fig:polH} 
Top : Longitudinal polarizabilities of H$_n$ chains (per hydrogen atom, in
${a_0}^3$) as a function of length $n$, for an alternation of 2 and 3
$a_0$ bond length in various calculations; we present our own LDA,
SIC, \JMcommNEW{standard SIC-Slater} and GSlat results 
(open symbols)
in comparison with calculations from other groups:
%SIC~\protect\cite{Ruz08}, LDA~\protect\cite{Ruz08},
\JMcommNEW{standard} SIC-KLI~\protect\cite{Kor08}, \JMcommNEW{standard}
SIC-OEP~\protect\cite{Kor08},
\JMcommNEW{Loc. SIC-KLI~\protect\cite{Pem08}}
and quantum
chemistry MP4~\protect\cite{Cha95}. 
Bottom: Transverse polarizabilities of H chains (our own results only). 
}
\end{figure}

Our calculations on polarizabilities in hydrogen chains
are presented in Fig.~\ref{fig:polH} and are compared with previous
results \cite{Pem08,Kor08,Cha95}.
%
%% \footnote{JM2o : Fig. 1 is not very readable. Any idea to improve it
%%   ?\\
%% \PGRcomm{PGR2JM: I have printed the page and found the figure
%%   acceptable,
%% although not ideal. I would leave it.}}
%
For sake of a fair comparison, we have taken care
of using the same
%(regularizing, in the case of Hydrogen atom)
pseudopotential as used in former calculations because we
found that using different pseudopotentials leads to different
absolute values of the \EScommNEW{highly sensitive} polarizabilities. 
\EScommNEW{This point may look  surprising especially in the case of
hydrogen for which inserting a pseudopotential is a matter of
practical convenience for regularizing grid representations. However, as
the grid size is optimized to the regularizing core, different core
widths lead to slightly different finite representations of the
wave functions (having of course the same energy). This may deliver
slighly different values of polarizabilities.} 
This has to be kept
in mind when comparing with MP4 results which are computed in a
different fashion.  But the comparison between the various DFT
approaches stays on safe grounds. 

\MDcommNEW{We first mention} that our
LDA and SIC calculations perfectly match previously published results
\cite{Ruz08} \MDcommNEW{(not shown in the figure)}. Furthermore the
trends are rather systematic: LDA 
strongly overestimates polarizabilities, as expected, and SIC
comes much closer to MP4 results. Still the
most relevant comparison, in our opinion, is that between
SIC and approximations thereof, because of the pseudopotential effects
mentioned above.
%\JMcommNEW{Results are reported in Fig.~\ref{fig:polH}.}

There are several interesting points showing up from
this comparison.
First, while \JMcommNEW{standard} SIC-Slater (open circles) and
\JMcommNEW{standard} SIC-KLI (full 
circles) calculations lead to results of varying quality, we see that
GSlat (open squares), even if 
not perfectly matching exact SIC, leads to overall acceptable results
showing more regular and realistic trends.
%% \JMcommNEW{
%% This result is due to the fact that the localized orbitals that enter
%% in the GSlat potential create higher barriers in the potential
%% \cite{Pem08}.  
%% }
However the agreement in
absolute values tends to degrade with increasing chain length,
in relative values (of GSlat compared to SIC) : from +8$\%$ to +25$\%$
from H$_4$ to H$_{12}$.  
\JMcommNEW{
Hence one can wonder whether less dramatic approximations to the
Generalized SIC-OEP formalism would improve the results.
An obvious next step is to use the \MDcommNEW{Generalized SIC-KLI}
correction instead of GSlat. 
We show in Fig.~\ref{fig:polH} the Loc. SIC-KLI results
obtained by \cite{Pem08} with the addition of the
approximate Generalized SIC-KLI term as given in
Eq.~(\ref{eq:pot_OEP5}). Mind that 
%% , as discussed at
%% the end of in Sec.~\ref{sec:formalism}, this approximation is well
%% established for symmetrical systems (as are H chains). Note also that
these results used another localization criterion than the symmetry
condition. As seen in the figure, perfect agreement with SIC is
achieved \MDcommNEW{(compare large open triangles with black
diamonds). This thus validates {\it a posteriori} the assumption done
to obtain (\ref{eq:pot_OEP5}).
%, that is the fact that the off-diagonal
%terms (\ref{eq:A_ij}) are negligible. 
We however recall that
Eq.~(\ref{eq:pot_OEP5}) has no robust fundation. The Loc. SIC-KLI results
nevertheless demonstrate the promising}
possibility to use a numerically less costly localization
criterion than the symmetry condition.
\MDcommNEW{Indeed, the equivalent GSlat in~\cite{Pem08} perfectly
agrees with ours.} 
% We nevertheless mentioned that we have
%extensively studied the numerical performances of the symmetry
%condition; our investigations will be reported in a forthcoming
%paper.
}

%Note, however, that the assumption of symmetry is major premisse to
%obtain the expression (\ref{eq:pot_OEP6}) and
%thus, this approximation \PGRcommNEW{is not guaranteed}
%to work systematically for asymmetrical systems.
\MDcommNEW{
The mechanism invoked to explain the better performance of SIC-KLI
(standard and generalized) on polarizabilities, over mere LDA or
standard SIC-Slater, \JMcommNEW{is that the SIC-KLI correction, or the
so-called response part of
the exchange-correlation potential defined in~\cite{Gis99a}, 
produces a counter-field effect~\cite{Gis99a}~:
%is the addition of the SIC-KLI correction, or the so-called response part of
%the exchange-correlation potential defined in~\cite{Gis99a}.
%The effect of this correction is actually twofold~: {\it i)} without an
%electric field, potential barriers between the H$_2$ units are higher
%than in LDA; {\it ii)}
with an external field, the exchange-correlation
potential behaves globally against it. 
%Both properties are of course
%closely linked and constitute what is called the counter-field
%effect~\cite{Gis99a}~: 
Electronic motion is hindered and
polarizabilities are thus reduced compared with a LDA
treatment. The counter-field effect is all the more efficient when one
goes from SIC-KLI to OEP~\cite{Kue04,Kor08}
%, and even more so when SIC is
%included~\cite{Kor08} 
or in a Generalized SIC-KLI
treatment~\cite{Pem08}.} 
In GSlat, no response potential is
present, so no counter-field effect is expected here.
Indeed, the average 
GSlat exchange-correlation potential in the presence of an external field is not
opposite to the electric potential, as the SIC-KLI or the OEP ones are
%Still, the average GSlat potential  does not follow the external
%potential as strongly as in LDA or standard SIC-Slater, or can even be 
%constant over the whole H chain
(see e.g. bottom panels of Fig.~2
in~\cite{Pem08} which correspond to our GSlat but obtained with
another localization criterion than the symmetry condition).
Actually, the performance of our GSlat approximation stands in the
property that
%, without external field,
much higher barriers (than in standard SIC-Slater or SIC-KLI) between
H$_2$ units 
appear in the exchange-correlation potential,
\JMcommNEW{which hinder more the electronic motion.
This has to be linked to the localized character of the orbitals that
enter into its calculation \cite{Pem08}.}
 
%This has also to 
%be linked to the derivative discontinuity~\cite{Per92} which degrades
%when self-interaction error increases, or equivalently, which is
%restored when SIC is taken into account~\cite{Ruz08,Zha98,Mor03}.
}

\EScommNEW{It would be extremely interesting to also test 
\MDcommNEW{the Generalized SIC-KLI} and its various approximate forms}
in less symmetrical systems. 
% for which they are {\it a priori}
%not necessarily adapted, in particular the simple expression
%Eq.~(\ref{eq:pot_OEP5}).
%\MDcommNEW{Indeed, we have checked in a few test cases that when an
%asymmetrical potential is involved, some of the off-diagonal terms
%$A_{ij}$, $i\neq j$, are of the \emph{same} order of magnitude as that
%of the $A_{ii}$, while they are assumed to dominate the sum
%(\ref{eq:A_ii&A_ij}) to get the approximate expression
%(\ref{eq:pot_OEP5}) used in~\cite{Pem08}. It also seems that in
%%symmetrical systems, this property still holds and is not
%accidental.}
\MDcommNEW{This calls for a systematic study which will be
reported in a forthcoming paper. Keeping this in mind,}
we will nevertheless present in the following 
a few examples of applications to less symmetrical systems in
the case of GSlat in comparison to exact SIC.

%The mechanism to explain the OEP values relies
%on the counter-field effect which is connected to the degree of
%localization of orbitals. A full detailed analysis of this mechanism in
%GSlat goes beyond the scope of this paper but we can already argue
%that the localization observed in the orbital set used to construct
%the GSlat Hamiltonian certainly goes in the right direction.

\EScommNEW{Before doing so, let us remark that}
the case of full \JMcommNEW{standard} OEP results of \cite{Kor08}
(close squares) deserve a special comment. Indeed, up to 
fluctuations, the obtained results perfectly match the MP4 ones (stars),
and thus somewhat differ from SIC ones. There is here a matter of
interpretation in the sense that, if the aim is to match MP4 results,
the results are perfect. But if it is to find a good approximation to
SIC then the agreement is not dramatically better than
GSlat ones (relative values of OEP compared to SIC : $-6\%$ to $-13\%$
from H$_4$ to H$_{12}$) 
\JMcomm{and is worse than the Loc. SIC-KLI ones}. 
Moreover GSlat \JMcommNEW{and  Loc. SIC-KLI} calculations seen as
approximations to full Generalized SIC-OEP are cheaper schemes, easily
applicable to much more complex systems than hydrogen chains.

Note finally that Fig. \ref{fig:polH} also shows the transverse
polarizability for those hydrogen chains. In that case, the GSlat
approximation reproduces perfectly the exact SIC results,
\JMcommNEW{as expected.}
 
%\MDcommNEW{This was expected since the counter-field effect should be
%absent for such an observable in H chains.} 

Now that the capability of GSlat on hydrogen chains has been
checked, we deeper analyze the properties of the system in a
small chain, namely H$_4$.
\begin{figure}[htbp]
\begin{center}
\epsfig{width=0.7\linewidth,file=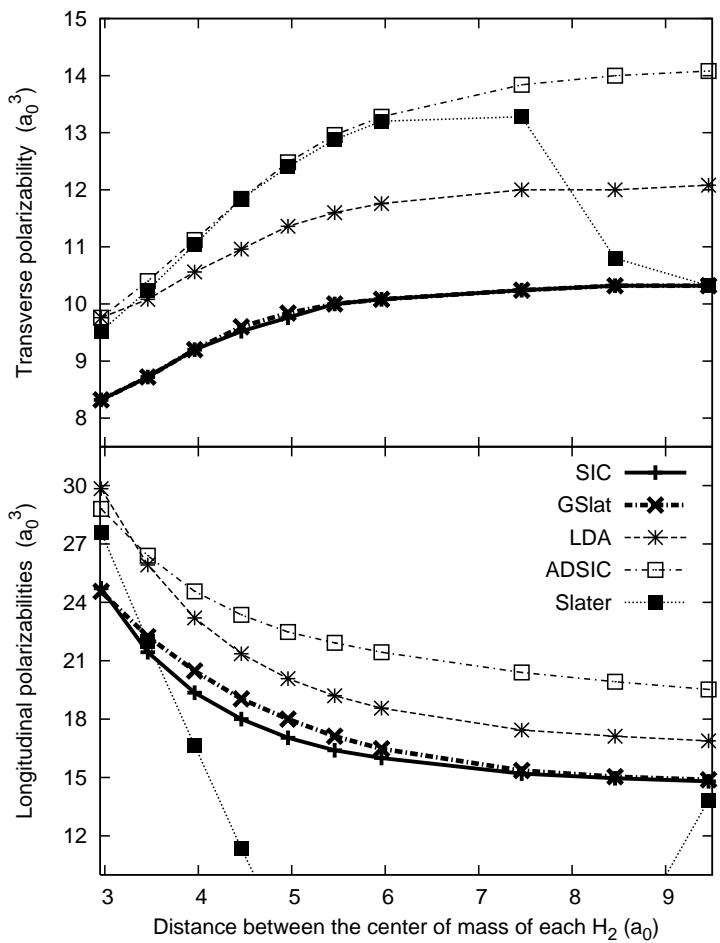}
\caption{\label{fig:polH4} 
\MDcomm{
Longitudinal (top) and tranverse (bottom) polarizabilities of H$_4$
chains, according to the H$_2$-H$_2$ center of mass distance,
for various SIC schemes as indicated.
}
}
\end{center}
\end{figure}
We present in Fig. \ref{fig:polH4} the values of the (longitudinal and
transverse) polarizabilities of H$_4$ chains, according to various
H$_2$-H$_2$ center of mass distances. Here we used the experimental
value of the H$_2$ bond length, that is 1.46 $a_0$. The data labeled
``SIC'' constitutes our benchmark. LDA (stars) overestimates polarizabilities which was
expected on the ground that LDA has a tendency to overmetallize
bonding.  The simplified ADSIC (open squares) scheme \cite{Leg02}
gives in general rather poor results.  Mind that ADSIC nevertheless
allows a fair reproduction of bonding properties in poly-acetylene
\cite{Cio05}. It obviously fails in the case of the more sensitive
polarizability.
GSlat (crosses) in turn reproduces very well the
exact SIC tendencies, while the standard SIC-Slater (full squares) is
completely wrong for intermediate intermolecular distances. 
This mismatch is correlated to a similar failure of standard
SIC-Slater in the potential energy surface at intermediate distances
\cite{Mes08-1}. And both failures can be tracked back to 
delocalization effects of the orbitals at critical configurations.  
%\footnote{JM2o : how going further in the explanation ? Say that we
%  checked this in our 1D model ?\\
%\PGRcomm{PGR2o: I would only mention that if we have published
%  citation for that.}}
GSlat allows to keep the wave functions entering the Hamiltonian 
localized and thus performs much better.
\JMcomm{We checked that \JMcommNEW{standard} SIC-KLI (not shown in the figure) does not
  cure this mismatch.} 
For large intermolecular distances, standard SIC-Slater and
GSlat results come close to each other because there
remain two separated H$_2$ molecules, which have each only one
electron in each spin subspace.

\subsection{The C$_2$ dimer and the C$_4$ chain}

Another interesting case is provided by small carbon chains whose
electrical excitation  properties
%\footnote{{PGR: That probably means
%``electrical excitation properties''. If yes it should be said so.}}
are well studied \cite{Bia02,Ord98,Wat92}.
%\footnote{Add citations :
%berkus, yabanabertsch,experiments,koutecky ?}. 
We consider here two examples, namely the C$_2$ dimer and the C$_4$
chain.  Since, to the best of our knowledge, there exist no experimental
results for the polarizabilities of those systems, our aim is to
compare various theoretical approaches using exact SIC as a benchmark.
We recall that we already demonstrated the quality of GSlat
in the case of a single carbon atom in \cite{Mes08-1} for
exchange only calculations. In exchange correlations calculations, the
C atom polarizability is again well reproduced by GSlat.
\JMcomm{As the electronic cloud in the C atom is slightly axially
deformed because of the single occupation of $2p_x$ and $2p_y$
orbitals (while the $2p_z$ orbital is unoccupied), one measures different
polarizabilities along the $x,y$ axes and along the $z$ axis.}
To put the subsequent results on C molecules into
perspective, we quote here briefly the polarizations for the C atom~:
along $z$ axis, $\alpha_z=10.40$ ${a_0}^3$ for both SIC and GSlat,
along $x,y$ axes, $\alpha_{x,y}= 11.52$ ${a_0}^3$ for GSlat
and 11.76 ${a_0}^3$ for SIC.  

%\footnote{ES2JM: is that true ? JM : yes.  
%\\
%\footnote{{PGR: Did we
%publish that? If yes, cite. If no, we should provide the numbers here.}}.
%

%% \begin{figure}[htbp]
%% \centerline{\includegraphics[width=5.5cm,angle=-90]{C2pol-d=2.5.eps}}
%% \caption{\label{fig:C2pol} 
%% Polarizabilities of the C$_2$ molecule calculated in various SIC
%% schemes. Horizontal lines emphasize the SIC benchmark
%% values and ease the comparison with the other results.
%% }
%% \end{figure}

%% \begin{figure}[htbp]
%% \centerline{\includegraphics[width=5.5cm,angle=-90]{C4pol.eps}}
%% \caption{\label{fig:C4pol} 
%% Polarizabilities of the linear C$_4$ molecule calculated in various
%% SIC schemes. 
%% }
%% \end{figure}

\begin{figure}[htbp]
\begin{center}
\includegraphics[width=\linewidth]{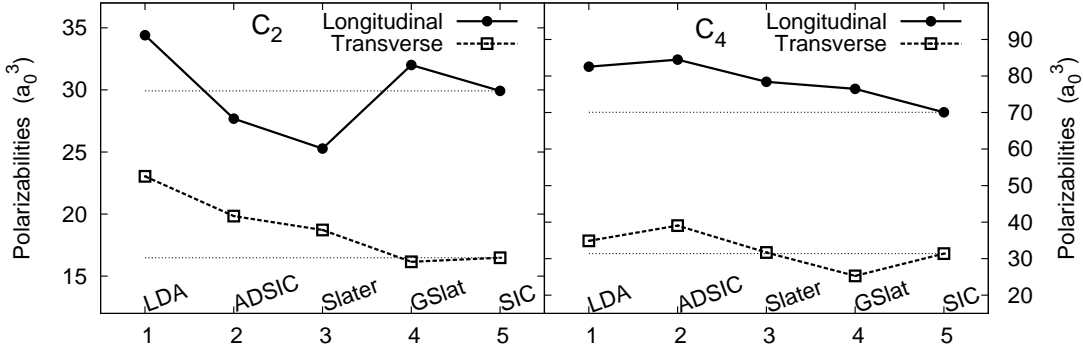}
\caption{\label{fig:C2-4pol} 
\MDcomm{
Transverse and longitudinal polarizabilities of the C$_2$ molecule
(left) and the C$_4$ chain (right), calculated in various SIC
schemes. Horizontal lines emphasize the SIC benchmark values and ease
the comparison with the other results.
}
}
\end{center}
\end{figure}

In Fig.~\ref{fig:C2-4pol}, we present the
longitudinal and transverse polarizabilities for C$_2$ and C$_4$
calculated in various approximations.  
{The calculations of \cite{Bia02,Abd02} yield comparable values.
In \cite{Bia02}, the longitudinal polarizability for C$_2$ is
$\alpha_\parallel=$ 
25 ${a_0}^3$ for the ab initio methods and 34 ${a_0}^3$ for LDA/GGA, while
the transverse one is $\alpha_\perp=$25 ${a_0}^3$ or 100 ${a_0}^3$
respectively, the latter value
being a strange exception. The results for C$_4$ are
$\alpha_\parallel=$ 92 or 94 ${a_0}^3$ and $\alpha_\perp=$
30 or 32 ${a_0}^3$. Our results are generally lower for
$\alpha_\perp$. It is worth noting that our calculations
differ in the employed functionals and pseudopotentials
which both can have a sensitive influence on the results.
Thus the comparison as a whole looks satisfying.
%\footnote{{PGR:
%That paragraph is a bit critical. We should check whether we go into
%that detail of comparison.}}.
To stay on the safe side, we concentrate on the comparison of
approaches within the same setup.
}

Fig.~\ref{fig:C2-4pol} shows again that GSlat
provides a very good approximation to exact 
SIC in C$_2$ and much better than any other approximation. The
situation is more mixed in the case of C$_4$.  {The values are
larger and the relative effects are smaller than for the C$_2$ dimer.
GSlat comes still closest to SIC for the longitudinal
mode, but standard SIC-Slater is competitive for the transverse mode.} Still,
when considering all cases together (C$_2$ and C$_4$, transverse and
longitudinal polarizabilities), it is clear that GSlat
provides a very good approximation (generally the best one) to
the exact SIC. 
%\footnote{ES: it would be great that we understand what is going on in
%C4 transverse ...!!!  JM : for me it is an accident (the maximum
%localization exceptionally gives a better result than the symmetry
%condition, and better than standard SIC-Slater) \\ {PGR: Note that
%the deviations from SIC are small for all approaches.  The remaining
%small differences should not be over-interpreted.}}

\subsection{Metal clusters}

As a final test case, we consider a small sodium cluster
representative of simple metallic systems. {We have chosen the
Na$_5$ cluster because it has a very soft electron cloud and is thus a
most critical test case amongst metallic particles \cite{Mun07c}. 
\begin{figure}[htbp]
\begin{center}
\includegraphics[width=0.8\linewidth]{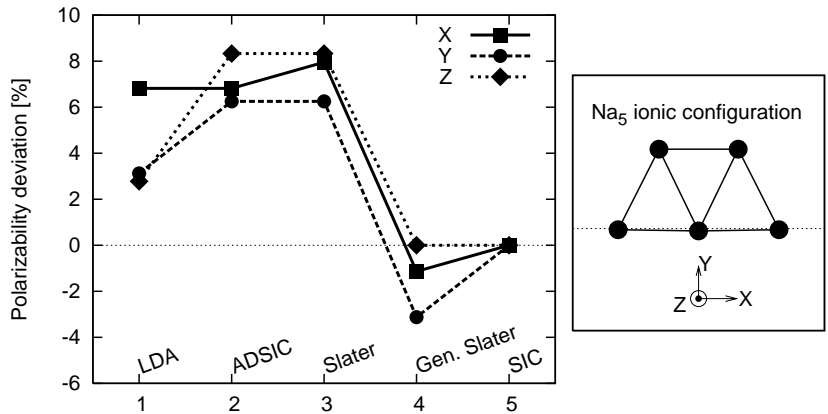}
\caption{\label{fig:na5pol} 
Polarizabilities of the Na$_5$ metal cluster (displayed in the right
panel), for various SIC schemes as indicated. Horizontal lines
emphasize the SIC benchmark values and ease the comparison with the
other results. 
}
\end{center}
\end{figure}
The cluster is planar (see structure in the insert of
Fig.~\ref{fig:na5pol}) which corresponds to a triaxial shape, and has 
accordingly different polarizabilities along the three major axes of
the system.}  Fig. \ref{fig:na5pol} shows the polarizabilities of the
Na$_5$.
{We obtain much larger absolute values of polarizabilities than in the
case of organic systems due to the metallic nature of bonding
(delocalization and lower binding). Not surprisingly, LDA performs
rather well, for sure better than in organic systems, {as is
to be} expected for a simple metallic system.} Even if LDA works very
well on those kind of systems, we still see a non negligible
difference with the benchmark SIC for the $X$ direction,
\JMcomm{about 7$\%$ for LDA}.
Again, GSlat stays {closest} to the
benchmark in all cases.

\section{Conclusion}

{We have tested a newly developed DFT-SIC scheme, called
Generalized SIC-Slater (GSlat), with respect to polarizability in chain
molecules and a soft metal cluster.  GSlat starts
from the Optimized Effective Potential (OEP) approach to SIC and
handles that in terms of two different sets of $N$ single-particle
wave functions. One set is taken for the solution of the OEP equations,
thus diagonal in energy and most likely delocalized. The other set is
used in setting up the SIC energy which becomes lowest for localized
wave functions. Both sets are connected by a unitary transformation
which leaves key features as, e.g, the total density invariant. Using
that double set allows to accommodate two conflicting demands, energy
diagonality versus locality. The unitary transformation is determined
by minimization of the SIC energy which leads to what is called the
symmetry condition, a key building block of the SIC equations.  The
localized character of the SIC optimizing set is well suited to
justify the steps from OEP to SIC-KLI and further to the SIC-Slater
approximation. Thus SIC-OEP with double-set representation and
subsequent SIC-Slater approximation leads to the GSlat
scheme. By virtue of the double-set technique, it has
\JMcomm{more flexibility}
%a wider range of applicability 
than standard SIC-KLI or SIC-Slater approximation.

As it is known that the polarizability in chain molecules is a
sensitive observable for DFT approaches, we have investigated the
performance of the new scheme with respect to polarizability in a
variety of critical test cases~: H chains which mimic the
electronic properties of polymers, 
%the T-shaped ground state of H$_4$,
the C$_2$ dimer, a C$_4$ chain, and Na$_5$ as a soft small metallic
particle.  
%
%The results demonstrate that the GSlat approximation
%performs very well in all test cases coming generally close to the
%values from full SIC which we use here as a benchmark.  It solves the
%pathologies of the standard SIC-Slater approximation which occur in
%critical (transitional) molecular configurations such that its
%performances depend very little on the kind of
%studied system and configuration (which is not the case in
%standard SIC-Slater or SIC-KLI).
%
The results demonstrate that the GSlat approximation comes generally
close to the values from exact SIC which we use here as a benchmark. It
solves the pathologies of the standard SIC-Slater approximation which
occur in critical (transitional) molecular configurations such that
its performances depend much less on the kind of studied system and
configuration (which is not the case in standard SIC-Slater or
SIC-KLI). For long H chains, some deviation is observed, which remains
reasonable compared to the other schemes
\JMcommNEW{and to the less important numerical cost of the GSlat scheme}. 
\JMcommNEW{
However addition of the SIC-KLI correction (coming from Generalized
SIC-OEP) allows to improve substancially the results. 
Nevertheless, 
\MDcommNEW{no strong argument justifies {\it a priori} the approximate form in
Eq.~(\ref{eq:pot_OEP5}), even} for spatially symmetrical systems. And
the case has yet to be tested for asymmetrical systems.
We here checked that GSlat performs fairly good as well in H chains than in less
symmetrical systems as C chains or Na clusters.
We finally checked that the improvement of the SIC-KLI correction over
GSlat is negligible when looking for other observables as energies.  
}
}

This work was supported,
%nr.  04670PG,
by  
%the CNRS Programme "Mat\'eriaux"   (CPR-ISMIR), 
%
%Institut Universitaire de France, 
Agence Nationale de la Recherche (ANR-06-BLAN-0319-02), 
the Deutsche Forschungsgemeinschaft (RE 322/10-1),
and the  Humboldt foundation.
% and a Gay-Lussac prize.

\end{document}